\makeatletter
\def\input@path{{./}{texdeps/}}
\IfFileExists{xstring.sty}{}{%
  \input{xstring.tex}\expandafter\def\csname ver@xstring.sty\endcsname{2099/01/01}}
\makeatother
\documentclass[sigconf, screen]{acmart}
\setcopyright{none}
\renewcommand\footnotetextcopyrightpermission[1]{}

\graphicspath{{fig/}}

\usepackage[linesnumbered,ruled,vlined,noend]{algorithm2e}
\usepackage{multirow}
\usepackage{adjustbox}
\usepackage{enumitem}
\usepackage{booktabs}

\definecolor{mylightgreen}{RGB}{0,160,89}

\SetCommentSty{commentstyle}

\newcommand{\myModel}{HeteroMoE}

\begin{document}

\title{Activation-Aware Balancing of Fused CPU and GPU Expert Execution for Offline MoE Inference}

\author{Borui Xu\texorpdfstring{\textsuperscript{1}}{}, Weihao Cui\texorpdfstring{\textsuperscript{1}}{}, Yao Chen\texorpdfstring{\textsuperscript{2}}{}, Weiguo Liu\texorpdfstring{\textsuperscript{3}}{}, Xuanhua Shi\texorpdfstring{\textsuperscript{2}}{}, Bingsheng He\texorpdfstring{\textsuperscript{1}}{}}
\affiliation{%
  \institution{\textsuperscript{1}National University of Singapore\quad \textsuperscript{2}Huazhong University of Science and Technology\quad \textsuperscript{3}Shandong University}
  \country{}}
\email{\{borui.xu, whcui, dcsheb\}@nus.edu.sg, \{chenyao\_cs, xhshi\}@hust.edu.cn, weiguo.liu@sdu.edu.cn}

\begin{abstract}
Frontier Mixture-of-Experts (MoE) models do not fit in single commodity GPU memory, so their routed experts stay in host memory and the CPU and GPU compute them together. The expert activation pattern of an MoE call, characterized by the number of activated experts and the tokens routed to each, affects the relative efficiency of CPU and GPU execution. In offline throughput-oriented inference, context length limits how many sequences fit in GPU memory, causing activation patterns and the optimal CPU--GPU partition to vary across workloads. Existing approaches sum fixed per-expert CPU costs, although a fused CPU call shifts between memory-bound and compute-bound execution as the activation changes. As a result, they often transfer too many experts to the GPU. To address these limitations, we present \myModel{}, an activation-aware expert scheduler that selects the CPU--GPU partition for each MoE call based on its live activation pattern and GPU cache state. Its roofline cost model captures the compute and memory limits of the fused CPU call and the PCIe weight transfers of the GPU branch, determining how many of the most heavily activated uncached experts to transfer. An engine with one synchronization per layer and optimized kernels runs this partition close to the measured hardware limits. On the H100 server, \myModel{} outperforms the strongest baseline in end-to-end throughput by 1.45--4.40$\times$ on DeepSeek-R1 and 1.51--3.70$\times$ on Kimi-K2. On the A6000 server, it is 1.16--3.86$\times$ faster on DeepSeek-R1.
\end{abstract}

\maketitle
\pagestyle{plain}
\renewcommand{\shortauthors}{Xu et al.}

\section{Introduction}

The rapid progress of large language models (LLMs)~\cite{gpt-3,touvron2023llama2openfoundation,opt,LLM_eva_survey,agent_survey} has made the Mixture-of-Experts (MoE) architecture~\cite{Switchtransformers,jiang2024mixtralexperts,mixtral-8x22b-v0.1,deepseekv2} a critical paradigm for expanding model capacity without proportional increases in computation. By activating only a small subset of experts per token, MoE models improve parameter efficiency while maintaining strong model quality~\cite{gemini25,deepseek-r1,qwen3}. In frontier MoE models such as DeepSeek-R1~\cite{deepseek-r1} and Kimi-K2~\cite{kimik2}, expert weights dominate the parameter footprint and cannot fit in a commodity GPU. Practical deployments therefore keep expert weights in host memory and execute the model using the CPU, GPU, and PCIe interconnect together.

In this paper, we target offline batch inference of frontier MoE models, including model evaluation~\cite{llamarena2024}, synthetic data generation, content generation~\cite{orca}, and text summarization~\cite{news_benchmarking}. In these workloads, requests are known in advance, and systems maximize throughput by batching as many sequences as GPU memory allows~\cite{deepspeed-inference,Pre-Gated-MoE,SiDA,MoE-Lightning,Klotski,MOE-Gen,MOE-Lens,ktransformers2024,Diff-MoE}. Since KV-cache usage grows with context length, workloads with different context lengths support different batch sizes and expert activation patterns. The activation workload of each expert affects whether CPU or GPU execution is more efficient, making the optimal CPU--GPU partition workload-dependent. Moreover, the change of expert activations and GPU-resident experts during inference require online partitioning.

The CPU and GPU execute their assigned experts concurrently, so the slower branch determines the MoE execution time. Since all expert weights reside in host memory, the CPU processes its assigned experts in a fused call, whose execution shifts between memory-bound and compute-bound regimes as activation patterns change. Offloading an expert to the GPU removes only that expert's share of this fused call, whereas every uncached expert costs a full PCIe weight transfer, however few tokens it serves. Therefore, the optimal CPU--GPU partition must \textbf{balance the fused CPU call against the PCIe transfers of the offloaded experts, using the live activation and cached experts of every MoE call}.

\begin{figure}[t]
  \centering
  \includegraphics[width=\columnwidth]{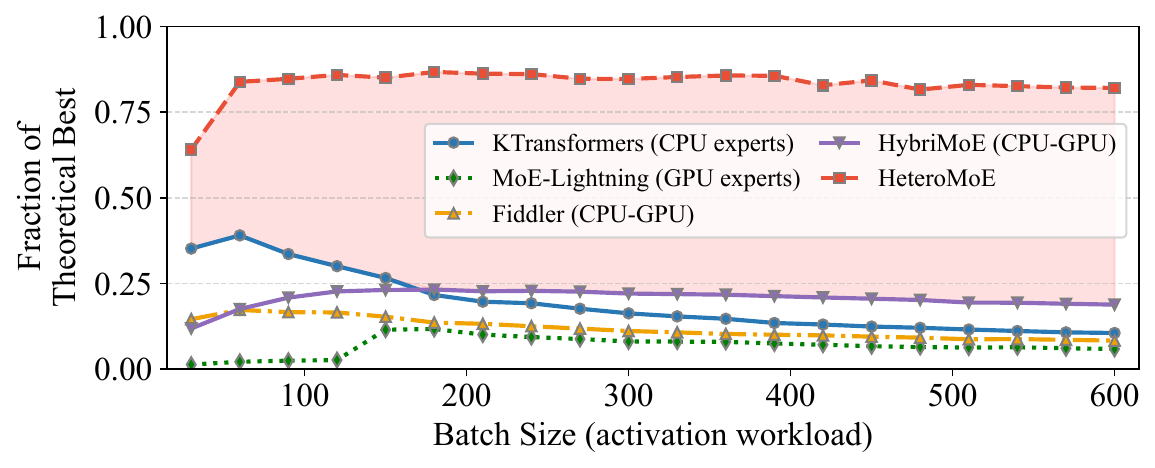}
  \caption{Throughput of MoE inference systems across expert activation loads on DeepSeek-R1 on H100, normalized to the theoretical best. The reference assumes the theoretically optimal CPU--GPU partition and an ideal expert cache.}
  \label{fig:batch_motivation}
  \vspace{-2em}
\end{figure}

However, existing systems fail to keep this balance across activation loads. Figure~\ref{fig:batch_motivation} shows their throughput relative to the theoretical best at each workload for DeepSeek-R1 with 32-token generation on an H100 server. We obtain this bound by giving every MoE call the roofline time of its best CPU--GPU partition, set by the measured hardware limits of host memory, PCIe, the CPU, and the GPU with an ideal expert cache of the same size, and keeping the measured time of all other work. KTransformers~\cite{ktransformers2024} computes all routed experts on the CPU, which suits light loads where each expert receives few tokens and reading its weights from host memory costs less than transferring them, but as the work per expert grows the CPU becomes the bottleneck, and its throughput falls from 35\% of the theoretical best at batch size 30 to 11\% at 600. GPU-only systems~\cite{MoE-Lightning,MOE-Gen,Klotski,MoE-Infinity,fmoe,Diff-MoE} transfer every non-resident expert over PCIe, which makes light activation loads transfer bound, and MoE-Lightning reaches at most 12\%. Fiddler and HybriMoE~\cite{Fiddler,HybriMoE} partition experts by live activation counts, but fixed costs per CPU expert overrate the fused CPU call, so they transfer too many experts and reach at most 23\%.

Our study reveals three insights that address these limitations. \textbf{First, the shortest MoE call balances the two branches, and the balance point moves with every call.} A lightly activated expert is cheaper on the CPU, which reads its weights from host memory, whereas a heavily activated one is cheaper on the GPU once its transfer is amortized. Since the batch size, the live routing, and the cached experts change the activation that each branch receives, a fixed device placement suits only one range of workloads, and the partition must be decided at every MoE call.

\textbf{Second, the activation is skewed, while every expert transfer costs the same.} The uncached experts of a layer have equal weight sizes, so each transfer takes the same time, whereas the CPU work it removes grows with the activation count. Transferring the hottest experts first therefore captures most of the benefit, and the scheduler only decides how many experts to transfer, which is cheap enough to evaluate the critical path at every call.

\textbf{Third, the fused CPU cost follows a roofline rather than a sum of expert costs.} The fused call is bounded by the slower of reading its expert weights from host memory and computing its tokens, which a sum of fixed costs per expert overrates. A roofline of these two limits estimates the fused CPU time accurately and avoids the excessive transfers that the additive estimate causes.

Based on these insights, we propose \myModel{}, an activation-aware CPU--GPU expert scheduler for high-throughput MoE inference. At every MoE call, its scheduler ranks the uncached experts by activation and picks the number of expert transfers with the shortest predicted critical path. An execution engine runs the CPU and GPU branches concurrently with a single GPU-to-host synchronization per layer, overlaps expert transfers with GPU computation, and employs optimized AVX-512 VNNI CPU kernels. Overall, we make the following contributions: 
\begin{itemize}[leftmargin=*,topsep=0.1em,itemsep=0.1em]
  \item We characterize how expert activation varies across offline workloads and decides whether the CPU or the GPU runs an expert faster, and we formulate the per-call CPU--GPU partition as a critical-path problem.
  \item We design an online, activation-aware and cache-aware scheduler that selects the expert partition of each decoding call in about 0.1 to 0.17~ms, using a calibrated cost model of the fused CPU call and the GPU branch.
  \item We build an expert engine that overlaps CPU computation, weight transfer, and GPU computation with one GPU-to-host synchronization per layer, together with an expert cache sized to the free GPU memory, AVX-512 VNNI CPU kernels, and a decode loop with little host overhead.
  \item On DeepSeek-R1, \myModel{} improves end-to-end throughput over the strongest baseline by 1.45 to 4.40$\times$ on an H100 server and by 1.16 to 3.86$\times$ on an A6000 server, and by 1.51 to 3.70$\times$ on Kimi-K2 on the H100 server.
\end{itemize}

\section{Background}

\subsection{Mixture-of-Experts Inference}

An MoE Transformer replaces the feed-forward network in each MoE block with a gate, a set of routed experts, and sometimes shared experts that process every token. For every token, the gate selects a fixed number of routed experts with the highest scores and produces routing weights. The selected experts process the token independently, after which their outputs are combined using the routing weights. Sparse routing increases model capacity without executing every parameter for every token.

Inference has two phases. Prefill processes all prompt tokens in parallel and usually produces dense expert activation. Decoding generates one token for each active sequence at every step, so its expert activation depends on the current batch and routing decisions. DeepSeek-R1 has 256 routed experts per MoE layer and selects eight per token, while Kimi-K2 has 384 routed experts and also uses top-8 routing. Both use the same expert shape, and their expert weights form most of the model footprint.

DeepSeek uses Multi-Head Latent Attention (MLA)~\cite{deepseekv2}, which stores a compact latent representation in the KV cache and reconstructs keys and values during attention. MLA reduces KV-cache memory compared with MHA~\cite{transformer} and GQA~\cite{GQA}, allowing attention and its cache to remain on the GPU for the workloads studied here. The KV cache still grows with batch size and generation length, leaving limited GPU memory for routed-expert weights.

\subsection{Heterogeneous CPU--GPU Expert Execution}

We consider a single-node server on which routed-expert weights reside in pinned host memory, while attention, the KV cache, gating, shared experts, and other dense operators remain on the GPU. At an MoE layer, the gate produces the routed expert IDs of the $T$ tokens of the call, where $T$ equals the active batch during decoding. Counting these IDs gives the activation count of each expert.

An active expert runs either on the GPU, after its weights cross PCIe unless they are already resident, or on the CPU, which reads them directly from host memory. The GPU adds the CPU outputs to its own, so moving an expert between devices changes only where it runs, not the token routing or the model semantics. Since the two devices run concurrently, a CPU--GPU partition trades fused CPU time against PCIe transfers and GPU kernels.

The CPU computes all its experts of a layer in one fused MoE call whose threads share the work. For each expert, the call gathers its routed tokens, computes the gate and up projections and then the down projection, and adds the outputs weighted by the routing weights. Because the call reuses each expert's weights across its routed tokens, the fused time follows the bytes of the distinct CPU experts when they receive few tokens and the token work when they receive many.

\section{Motivation and Characterization}
\label{sec:analysis}

We characterize how the expert activation varies across workloads and how it governs CPU--GPU expert partitioning.

\subsection{Activation Varies with Workload and Input}
\label{sec:batch_changes}

\begin{figure}[t]
  \centering
  \includegraphics[width=\columnwidth]{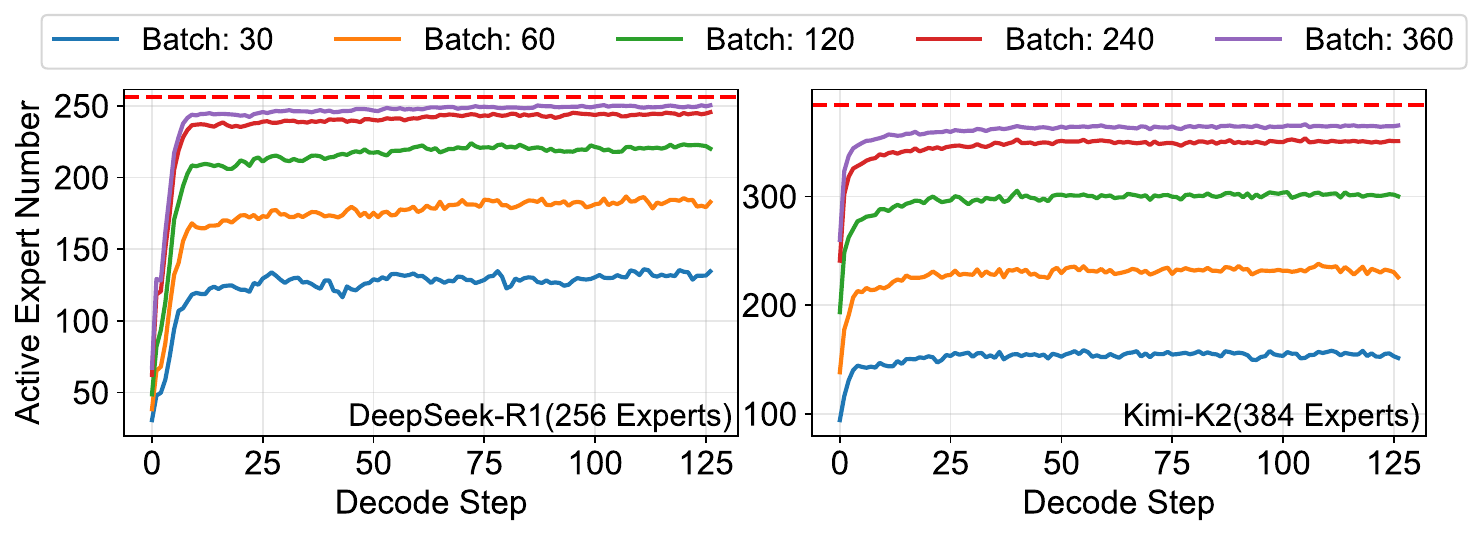}
  \caption{Average active experts per MoE layer of DeepSeek-R1 and Kimi-K2 on ChatGPT-prompts. Coverage rises over the first decoding steps and grows with the batch.}
  \label{fig:expert_act_per_step}
  \vspace{-1em}
\end{figure}

\begin{figure}[t]
  \centering
  \includegraphics[width=\columnwidth]{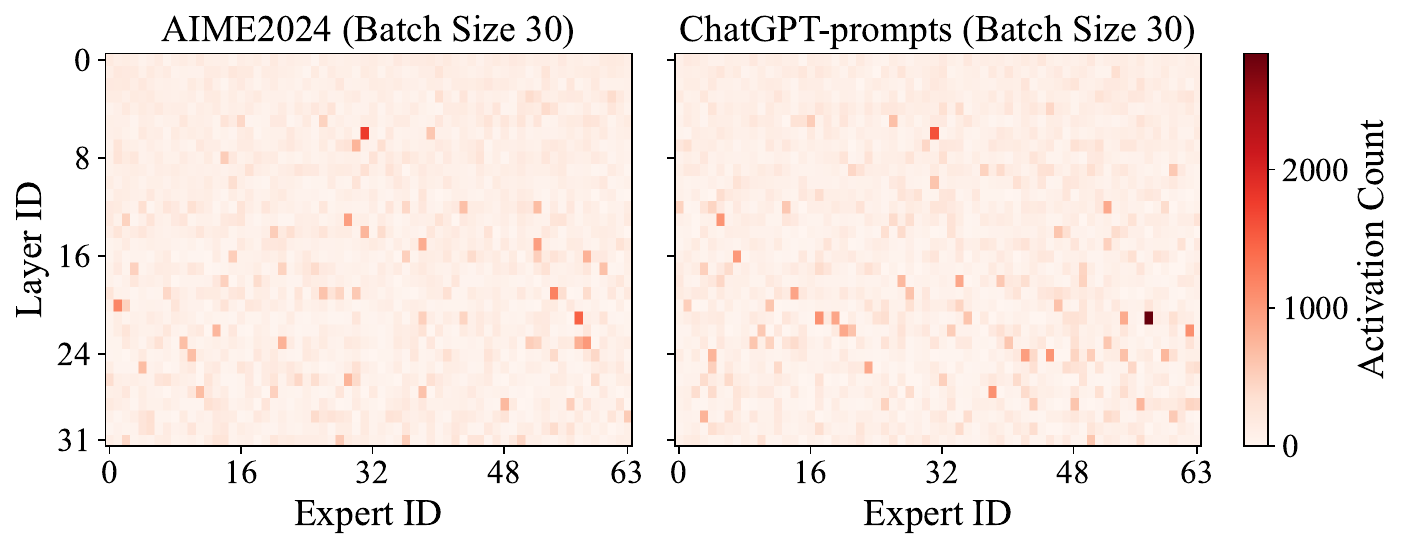}
  \caption{Local heatmaps of expert activation counts across datasets. Each dataset has its own hot experts.}
  \label{fig:expert_act_heatmap}
  \vspace{-1em}
\end{figure}

\textbf{Observation 1. Both the amount and the identity of expert activation change across workloads.} A longer sequence budget reserves more KV-cache memory under the same GPU memory limit, which reduces the feasible batch and the tokens of each MoE call. Figure~\ref{fig:expert_act_per_step} shows the average number of active experts per MoE layer of DeepSeek-R1 and Kimi-K2 on ChatGPT-prompts. Larger batches activate more experts, and once the coverage nears saturation, further batch growth mainly raises the activation count of each expert. The input also decides which experts are hot. At the same batch size of 30, the top-10 experts of AIME2024 and ChatGPT-prompts overlap by only about 13\% across layers, and those of four other datasets overlap with ChatGPT-prompts by 17 to 67\%, as Figure~\ref{fig:expert_act_heatmap} shows. The work per expert and the hot set therefore both change with the workload, so neither a fixed device placement nor a statically pinned hot set fits all workloads, and the partition must follow the live routing of each call.

\subsection{Skewed Activations with Short-Term Reuse}
\label{sec:skew_stable}

\begin{figure}[t]
  \centering
  \includegraphics[width=\columnwidth]{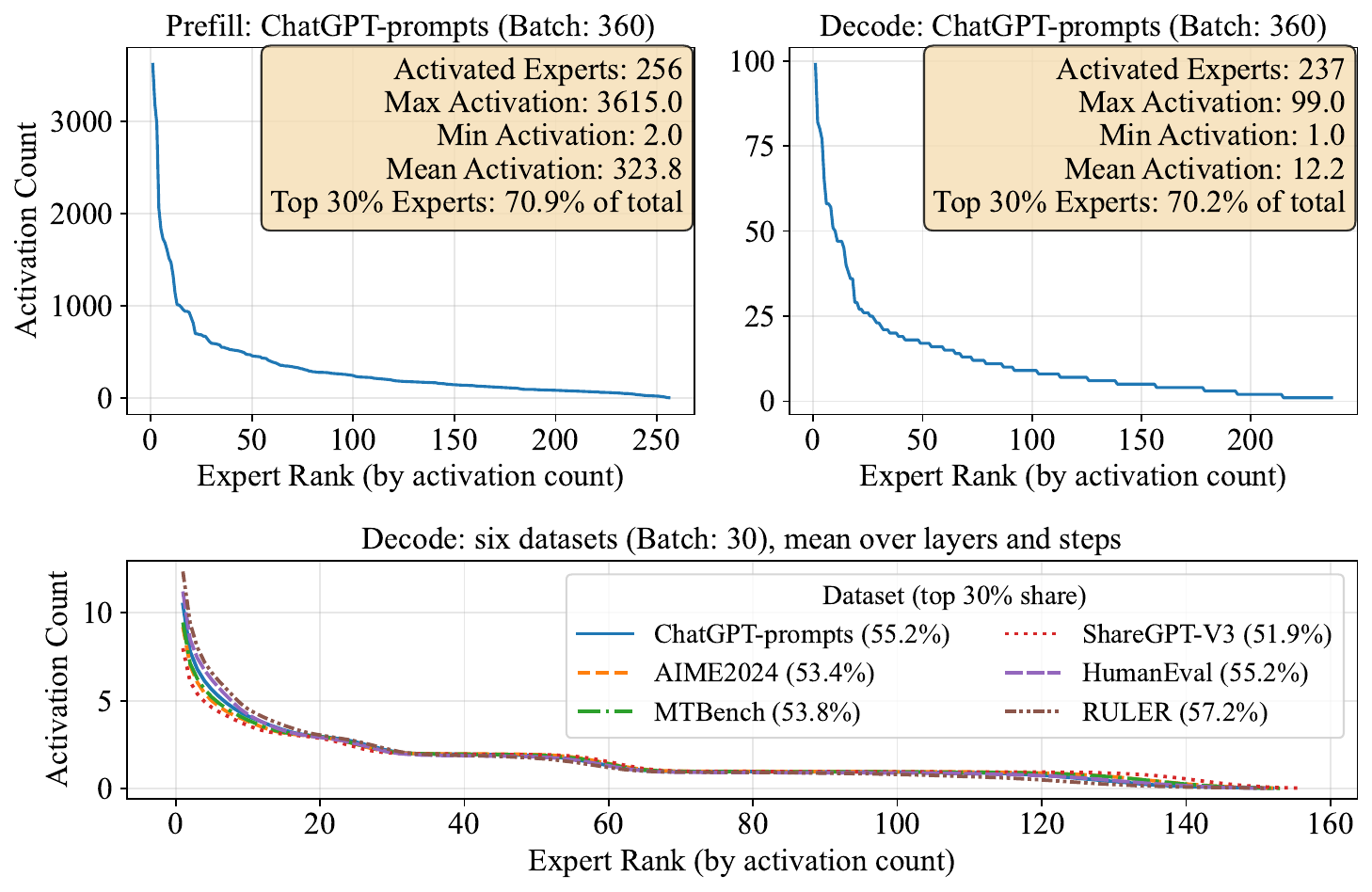}
  \caption{Ranked expert activation count distributions in prefill and decode phases (top) and across six datasets at batch size 30 (bottom). All exhibit a skewed distribution.}
  \label{fig:activation_skew}
  \vspace{-1.5em}
\end{figure}

\textbf{Observation 2. Activations are skewed, and hot experts recur across adjacent steps.} Figure~\ref{fig:activation_skew} ranks activation counts for one DeepSeek-R1 MoE layer on ChatGPT-prompts at batch size 360 during prefill and decoding. In both phases, the top 30\% of experts receive about 70\% of all routed activations. The bottom panel shows that decode activations stay skewed across six datasets from math, chat, code, and long-context tasks, where the top 30\% of experts receive 52 to 57\% of activations at batch size 30. This skew favors placing the most heavily activated experts on the GPU, because a transfer costs the same for every expert of a layer while the CPU work it removes grows with the activation count of the expert.

Hot experts also recur over short time spans. In our traces, more than 60\% of the active experts of a layer are reused between adjacent decoding steps even at batch size 30, and reuse increases with batch size. A score that decays the past activation counts of each expert therefore predicts the experts of the next step well. Recent routing thus guides caching and prefetching, while the live activation and the current cache content decide how many and which experts run on the GPU at every call.

\subsection{Balancing the CPU--GPU Critical Path}
\label{sec:critical_path}

\begin{figure}[t]
  \centering
  \includegraphics[width=\columnwidth]{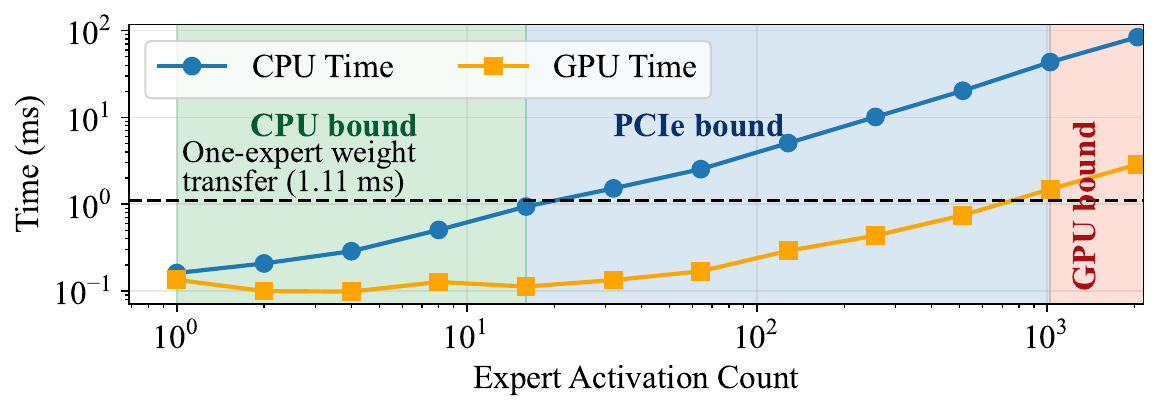}
  \caption{Per-expert CPU and GPU time on A6000. The CPU is faster at low activation counts and the GPU at high counts, and the dashed line marks the PCIe transfer of one expert.}
  \label{fig:flops_time}
\end{figure}

\textbf{Observation 3. The fastest partition balances the concurrent CPU and GPU+PCIe paths.} Figure~\ref{fig:flops_time} measures per-expert CPU and GPU costs for DeepSeek-R1 on A6000. The dashed line marks the 1.11~ms PCIe transfer of one expert's weights. CPU and GPU costs scale differently with an expert's activation count. The CPU is faster at low counts because it avoids weight transfer, whereas the GPU becomes preferable at high counts once transfer is amortized. As activation counts are skewed, experts in the same layer can therefore prefer different devices.

\begin{figure}[t]
  \centering
  \includegraphics[width=\columnwidth]{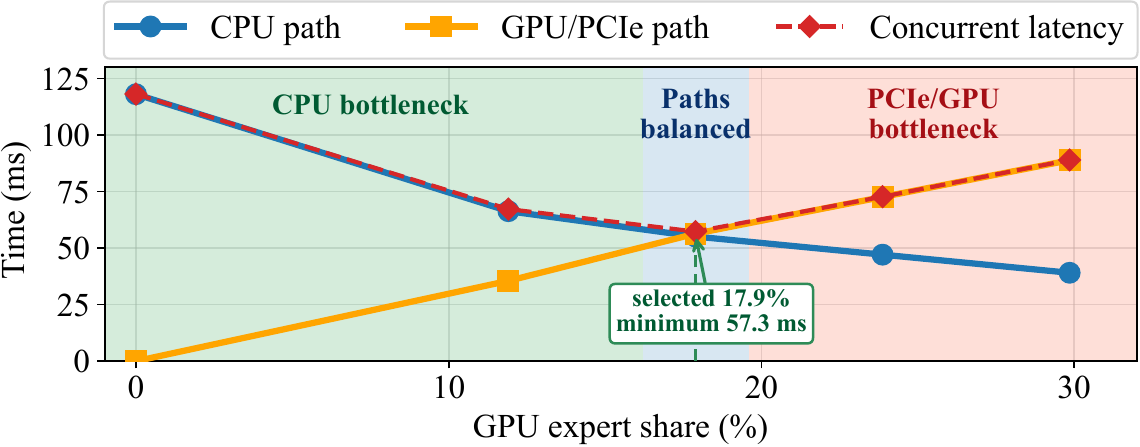}
  \caption{Critical-path time versus GPU share for DeepSeek-R1 on A6000 at batch 600. The layer time is lowest where the CPU and GPU+PCIe branches balance.}
  \label{fig:static_scheduling}
  \vspace{-1.5em}
\end{figure}

At layer scale, the CPU and GPU+PCIe branches overlap, so latency is determined by the slower branch. Figure~\ref{fig:static_scheduling} illustrates this tradeoff for DeepSeek-R1 on A6000 using ShareGPT-V3 at batch 600. A small GPU share leaves execution CPU-bound, whereas a large share makes the GPU+PCIe branch the bottleneck. Latency is minimized when the two branches are balanced, so effective partitioning must balance the complete CPU--PCIe--GPU path rather than either device alone.

\subsection{Fused CPU Time Follows Bytes and Tokens}
\label{sec:fused_cpu}

\begin{figure}[t]
  \centering
  \includegraphics[width=\columnwidth]{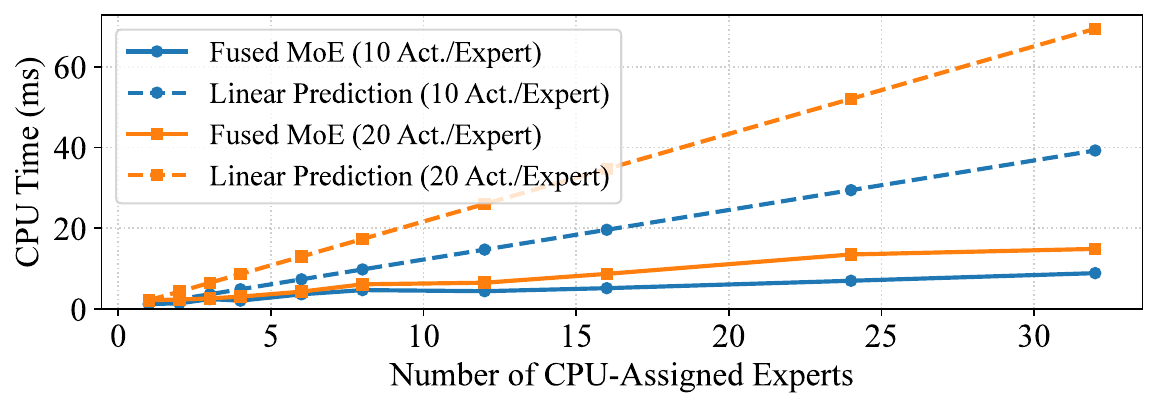}
  \caption{Measured fused CPU MoE latency versus additive linear predictions under two uniform expert loads. The additive predictions overrate the fused time.}
  \label{fig:cpu_fused_scaling}
  \vspace{-2em}
\end{figure}

\textbf{Observation 4. Fused CPU execution shifts between memory and compute bounds.} Fiddler and HybriMoE~\cite{Fiddler,HybriMoE} charge each CPU expert a fixed per-expert cost and sum these costs. Real CPU backends instead fuse all CPU-assigned experts of a layer into one call that shares its threads and synchronization. As Figure~\ref{fig:cpu_fused_scaling} shows, the fused CPU time does not grow linearly with the number of assigned experts, even under uniform loads. On our A6000 server, for example, a ShareGPT-V3 decoding call at batch 600 runs about 200 active experts on the CPU. An additive model calibrated from single-expert runs predicts 328~ms for these experts, whereas their fused call takes only 68.7~ms. Such a model overrates the CPU cost of moderately loaded experts and sends too many experts to the GPU, as Section~\ref{sec:policy_comparison} shows.

The fused call is instead bounded by two resources. With few tokens per expert, it reads the weights of every CPU expert from host memory and is bound by DRAM bandwidth. On the H100 server, the CPU branch reads expert weights at about 300~GB/s at batch sizes 30 and 80, close to the 350~GB/s that the same CPU threads reach in a read benchmark. With many tokens per expert, the token computation dominates. A model of the CPU branch must therefore count distinct expert bytes and token work rather than add independent per-expert costs.

\section{\myModel{} Design}
\label{sec:design}

\subsection{Design Requirements and Overview}

Section~\ref{sec:analysis} identifies four requirements for high-throughput systems. First, the partition must follow the live expert activation and the GPU cache content at every MoE call. Second, it must decide in the same step how many experts use the GPU and which experts fill that share. Third, its cost estimate must capture the memory and compute bounds of the fused CPU call and the transfer cost of every uncached expert, which shares host memory with the CPU branch. Fourth, the backend must run the two branches concurrently, so that the slower branch, and not host overhead, sets the layer time.

\begin{figure}[t]
  \centering
  \includegraphics[width=\columnwidth]{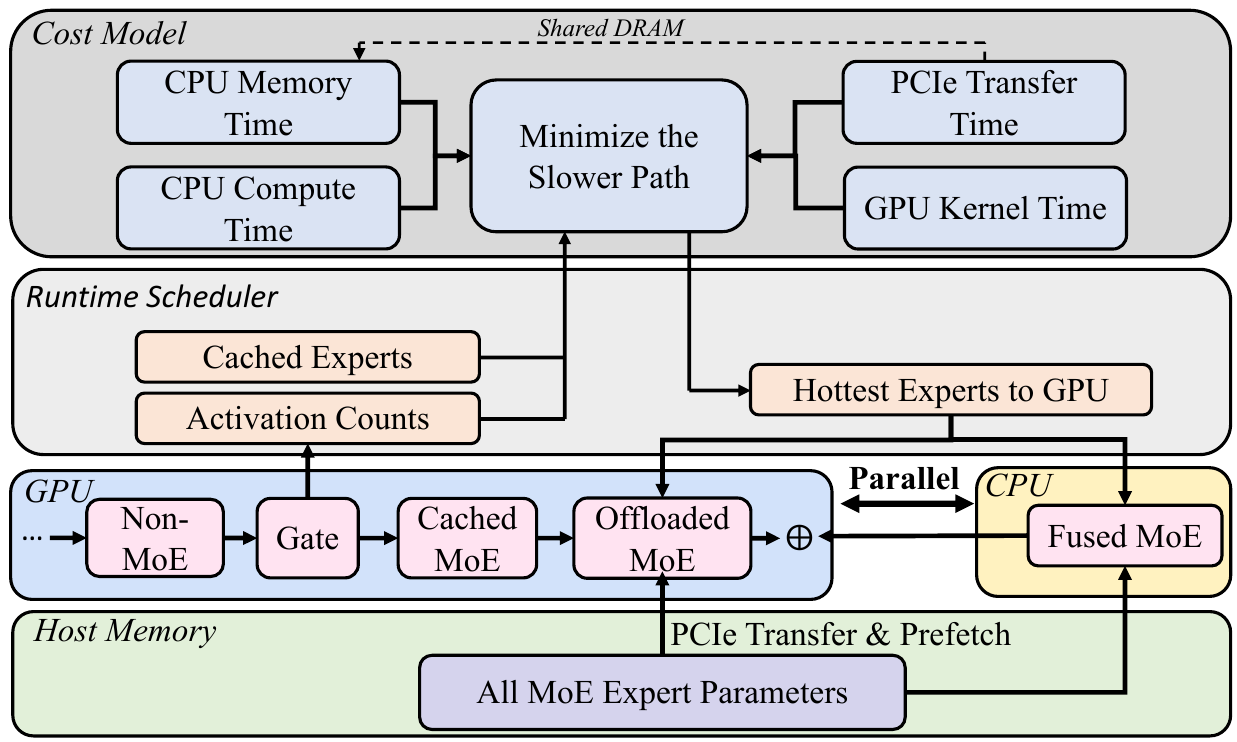}
  \caption{Overview of \myModel{}. Per MoE layer, the runtime scheduler sends cached and hottest uncached experts to the GPU and the rest to the fused CPU kernel.}
  \label{fig:sys_overview}
  \vspace{-1em}
\end{figure}

To satisfy these requirements, we propose \myModel{}, whose architecture Figure~\ref{fig:sys_overview} shows. At each MoE layer, the runtime scheduler counts the activation of every expert from the routed expert IDs and identifies the cached experts. It then picks how many of the hottest uncached experts to transfer so that the slower of the two branches finishes earliest. The cached experts run in the cached MoE block and the transferred ones in the offloaded MoE block on the GPU, while the remaining experts run in the fused MoE block on the CPU. The two branches run in parallel and synchronize only to merge their outputs, and the expert cache and prefetching keep frequently used experts on the GPU.

\subsection{Activation-Aware Expert Scheduler}
\label{sec:online_partitioning}

At each MoE call, the runtime scheduler determines how many and which active experts to execute on the GPU. Table~\ref{tab:notation} summarizes the variables used by the scheduler. Let $c_e$ denote the activation count of expert $e$, and $\mathcal E_{\mathrm{gpu}}$ the set of active experts whose weights already reside on the GPU. These experts can execute on the GPU without weight transfers. The other $N$ active experts are sorted by activation count in descending order as $e_1,\dots,e_N$. The scheduler transfers the first $m$ of them to the GPU, and the remaining $N-m$ experts run on the CPU. The cost model estimates the CPU time and the GPU time of the call as functions of $m$.

\begin{table}[t]
\centering
\caption{Variables of the scheduler.}
\label{tab:notation}
\begin{adjustbox}{max width=\columnwidth}
\begin{tabular}{@{}ll@{}}
\toprule
Symbol & Meaning \\
\midrule
$c_e$ & activation count of expert $e$ \\
$\mathcal E_{\mathrm{gpu}}$ & active experts already on the GPU \\
$N$ & number of the other active experts, sorted by $c_e$ as $e_1,\dots,e_N$ \\
$m$ & number of transferred experts, which are the first $m$ \\
$W(m)$ & CPU token work, $\sum_{i>m}c_{e_i}$ \\
$t_{\mathrm{rd}}$ & time to read one expert weights from host memory \\
$t_{\mathrm{tok}}$ & CPU time per unit of token work \\
$\theta$ & share of $t_{\mathrm{rd}}$ that one transfer takes from the CPU \\
$t_{\mathrm{tr}}$ & time to transfer one expert to the GPU \\
$t_{\mathrm{ker}}(c)$ & GPU kernel time of an expert with $c$ tokens \\
\bottomrule
\end{tabular}
\end{adjustbox}
\vspace{-1em}
\end{table}

\subsubsection{CPU Cost Model}

\begin{figure}[t]
  \centering
  \includegraphics[width=\columnwidth]{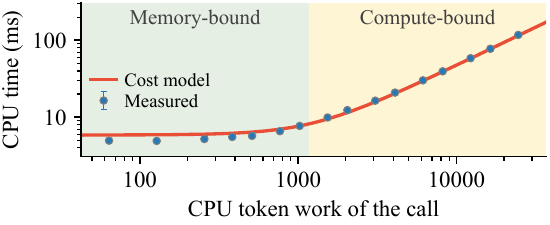}
  \caption{Measured and modeled fused CPU time for 64 experts under varying token loads on H100.}
  \label{fig:cpu_roofline}
  \vspace{-1.5em}
\end{figure}

Observation~4 shows that fused CPU execution is limited by either reading expert weights from host memory or computing its token workload $W(m)=\sum_{i>m}c_{e_i}$. We therefore model memory and compute costs separately:
\begin{equation}
  \label{eq:parts}
  t_{\mathrm{mem}}(m)=t_{\mathrm{rd}}\,(N-m+\theta m),\qquad t_{\mathrm{comp}}(m)=t_{\mathrm{tok}}\,W(m)
\end{equation}
The memory cost includes $t_{\mathrm{rd}}$ for each of the $N-m$ CPU experts and a small calibrated share $\theta\,t_{\mathrm{rd}}$ for each weight transfer, which reads the same host memory. The compute cost is proportional to the CPU token workload, with $t_{\mathrm{tok}}$ per unit. For $m<N$, CPU execution time is estimated as:
\begin{equation}
  \label{eq:tcpu}
  t_{\mathrm{cpu}}(m)=t^{0}_{\mathrm{cpu}}+\max\big(t_{\mathrm{mem}}(m),\,t_{\mathrm{comp}}(m)\big)
\end{equation}
where $t^{0}_{\mathrm{cpu}}$ is the overhead of a call for submitting the CPU task and exchanging its inputs and outputs with the GPU, and $t_{\mathrm{cpu}}(N)=0$. As Figure~\ref{fig:cpu_roofline} shows, the memory cost dominates at low token loads, while the compute cost dominates at high loads.

\subsubsection{GPU Cost Model}

Expert weight transfers run sequentially on the copy engine, each taking $t_{\mathrm{tr}}$, while the kernel time $t_{\mathrm{ker}}(c)$ grows linearly with the expert's token count $c$. Each transferred expert executes after its weights arrive, while GPU-resident experts execute between transfers. We model GPU time using two parts, the completion time of the last transfer and its kernel, and the total kernel time:
\begin{equation}
  \label{eq:gparts}
  \begin{aligned}
  t_{\mathrm{xfer}}(m)&=t_{\mathrm{tr}}\,m+t_{\mathrm{ker}}(c_{e_m}),\\
  t_{\mathrm{kern}}(m)&=\textstyle\sum_{e\in\mathcal E_{\mathrm{gpu}}}t_{\mathrm{ker}}(c_e)+\sum_{i\le m}t_{\mathrm{ker}}(c_{e_i})
  \end{aligned}
\end{equation}
where $t_{\mathrm{xfer}}(0)=0$. The GPU time is estimated as:
\begin{equation}
  \label{eq:tgpu}
  t_{\mathrm{gpu}}(m)=t^{0}_{\mathrm{gpu}}+\max\big(t_{\mathrm{xfer}}(m),\,t_{\mathrm{kern}}(m)\big)
\end{equation}
where $t^{0}_{\mathrm{gpu}}$ is the overhead of a call for launching its kernels.

\myModel{} calibrates the constants of both models once per server and model with short calibration runs. A prefill benchmark at two prompt lengths measures the transfer time, the GPU kernel times, and the CPU branch of large calls, and one decoding run per batch size sweeps the number of transferred experts across both sides of the balance point. A least-squares fit then sets all constants automatically. Sections~\ref{sec:model_accuracy} and~\ref{sec:overhead} report its accuracy and time.

\subsubsection{Partition Decision}

Since the two branches run concurrently, the scheduler picks:
\begin{equation}
  \label{eq:argmin}
  m^\ast=\arg\min_{0\le m\le N}\ \max\big(t_{\mathrm{cpu}}(m),\,t_{\mathrm{gpu}}(m)\big)
\end{equation}
For a given $m$, transferring the most frequently activated experts maximizes the token work offloaded to the GPU, since all experts in a layer have the same weight size. Together with the cached experts, the scheduler only needs to determine the expert transfer number $m$, while the activation counts determine which experts to transfer. Algorithm~\ref{alg:scheduler} summarizes this process. Prefix sums over sorted activation counts efficiently compute $W(m)$ and GPU kernel costs for all candidate values of $m$, avoiding exhaustive search over expert subsets. The scheduling overhead is only 0.10--0.17~ms per decoding call, as Section~\ref{sec:overhead} shows.

\begin{algorithm}[t]\small
\caption{Activation-Aware Expert Scheduling}
\label{alg:scheduler}
\KwIn{Routed expert IDs of the call; experts resident on the GPU}
\KwOut{Number $m^\ast$ of transferred experts}
$c_e\gets$ number of routed IDs of expert $e$\;
$\mathcal E_{\mathrm{gpu}}\gets$ active experts resident on the GPU\tcp*[r]{no transfer}
$e_1,\dots,e_N\gets$ other active experts by descending $c_e$\;
\For(\tcp*[f]{vectorized}){$m\gets0$ \KwTo $N$}{
  evaluate $\max\big(t_{\mathrm{cpu}}(m),\,t_{\mathrm{gpu}}(m)\big)$\;
}
$m^\ast\gets$ the $m$ with the smallest value\;
transfer $e_1,\dots,e_{m^\ast}$ and send $e_{m^\ast+1},\dots,e_N$ to the CPU\;
\KwRet{$m^\ast$}\;
\end{algorithm}

This decision adapts to both the cache and the activation. A warm cache moves work to the GPU without any transfer, so the scheduler needs fewer transfers to balance the two branches, whereas a cold cache shifts the balance point toward the CPU. Calls with different activations also receive very different partitions. For example, for DeepSeek-R1 on the H100 server, prefill calls with thousands of tokens transfer 57 to 195 of their 213 to 227 uncached experts, whereas decoding calls transfer 12 to 37.

\subsection{Critical-Path-Aware CPU--GPU Execution}
\label{sec:engine}

\begin{figure}[t]
  \centering
  \includegraphics[width=\columnwidth]{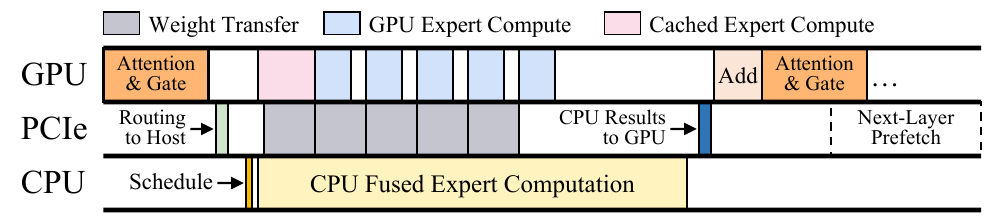}
  \caption{Pipelined CPU and GPU expert execution of one MoE layer. The CPU task starts right after scheduling, and each transferred expert runs once its weights arrive.}
  \label{fig:engine}
  \vspace{-1em}
\end{figure}

In Figure~\ref{fig:sys_overview}, the scheduler sends the CPU experts to the fused MoE block and the cached and transferred experts to the cached MoE and offloaded MoE blocks on the GPU. This partition minimizes the layer time only if the two branches execute concurrently and the host adds no delay between them. Figure~\ref{fig:engine} details the execution pipeline of one MoE layer.

At each MoE call, a single GPU-to-host synchronization makes the routed expert IDs and routing weights available in host memory. The scheduler first submits the fused CPU task, allowing the CPU branch with most active experts to start as early as possible. It then enqueues the GPU work without further host-side waits. On the GPU, cached experts are computed while uncached expert weights are transferred, and each transferred expert is computed as soon as its weights arrive. By overlapping weight transfers with GPU computation, this execution keeps the GPU branch close to the transfer time when transfers dominate.

After both branches complete, the GPU reads the CPU outputs from host memory and merges them with the GPU outputs using the original routing weights. The MoE execution time is therefore approximately the longer of the CPU branch and the overlapped GPU computation and PCIe transfers, plus the merge time. The scheduler aims to minimize this execution time. The next layer's attention starts immediately after the merge. During attention, PCIe and host memory would otherwise remain idle, so \myModel{} uses them to prefetch predicted experts for the next MoE layer, as described in Section~\ref{sec:cache}.

\subsection{Cache-Aware Expert Weight Management}
\label{sec:cache}

\begin{figure}[t]
  \centering
  \includegraphics[width=\columnwidth]{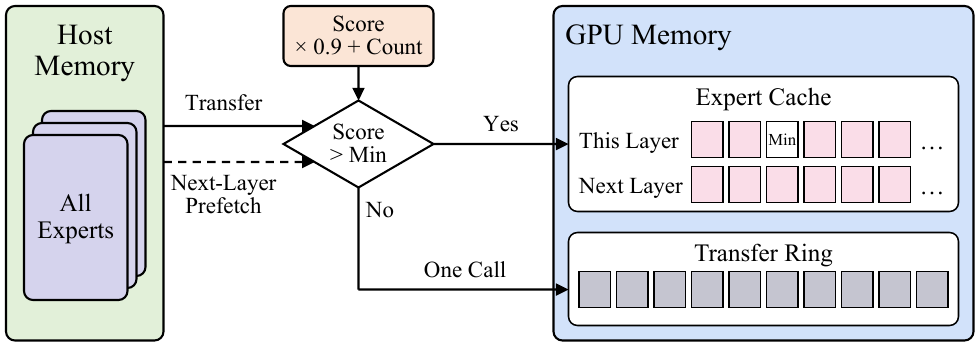}
  \caption{Expert cache and transfer ring. A transferred or prefetched expert replaces the lowest-scored cached expert if its score is higher; otherwise, it stays in the transfer ring for one call.}
  \label{fig:cache}
  \vspace{-1em}
\end{figure}

Figure~\ref{fig:sys_overview} distinguishes cached and offloaded experts by whether their weights reside on the GPU. After allocating memory for non-expert weights, the KV cache, attention tensors, and transfer buffers, \myModel{} uses the remaining GPU memory for per-layer expert caches. Observation~2 shows that hot experts tend to remain active across adjacent steps. However, the benefit of caching an expert depends on how many tokens it processes, rather than how recently it was used.

\myModel{} therefore manages the cache using recent activation counts. Each expert maintains a score updated at every call:
\begin{equation}
    s_e^{(t)} = 0.9 s_e^{(t-1)} + c_e^{(t)},
\end{equation}
where $c_e^{(t)}$ is the number of tokens routed to expert $e$ in call $t$. A transferred expert replaces the lowest-scored cached expert only if its score is higher; otherwise, it stays in a transfer slot for the current call. Figure~\ref{fig:cache} illustrates this placement.

The same scores also guide prefetching. While the host waits for the current CPU branch, \myModel{} transfers up to eight highest-scored uncached experts of the next MoE layer, as many as can finish before that layer's own transfers begin, so the prefetch overlaps the next layer's attention and gating. A prefetched expert enters the cache if its score exceeds the lowest cached score and otherwise stays staged in a transfer slot for the next call only. Algorithm~\ref{alg:scheduler} counts both cached and staged active experts in $\mathcal E_{\mathrm{gpu}}$, so a correct prediction removes a transfer from the next call.

The cache also affects CPU--GPU partitioning. Cached active experts always execute on the GPU without transfers, and Algorithm~\ref{alg:scheduler} accounts for their computation when partitioning the remaining experts. A warm cache thus moves more work to the GPU without transfer overhead. When routing changes, fewer active experts are cached, increasing GPU transfer costs. The scheduler then assigns more uncached experts to the CPU until the cache adapts.

\section{Implementation Optimizations}
\label{sec:impl}

\myModel{} is built on KTransformers~\cite{ktransformers2024}. Expert weights are quantized offline with K-quant~\cite{k_quant} into GGUF, the quantized weight format of llama.cpp, and kept in pinned host memory as a single GGML-formatted copy~\cite{ggml2023} that the CPU and GPU kernels share. The scheduler and the execution engine use CUDA streams and events, and the CPU kernels extend the CPU backend.

To realize the design in Section~\ref{sec:design}, our implementation must keep both execution branches close to their hardware limits while minimizing host-side overhead. In this secion, we introduce our implementation optimization.

\subsection{Execution Engine}
\label{sec:engine_impl}

Following the pipeline in Section~\ref{sec:engine}, the engine asynchronously copies the expert IDs and routing weights of each call to pinned host buffers and waits on a single CUDA event. This is the only GPU-to-host synchronization before the final wait for the CPU branch. The host then computes the partition, prepares the CPU expert IDs, and submits the CPU task. The GPU branch sorts token--expert pairs by expert, gathers their inputs once, executes the fused gate-and-up GEMM and down GEMM on contiguous token rows, and accumulates the outputs with a single weighted scatter-add. Experts operate directly on quantized weights, except in calls with thousands of tokens, where experts receiving at least 64 tokens use cuBLAS BF16 GEMMs with temporarily dequantized weights. Expert transfers use a ring of slots on a separate copy stream, where a slot is reused after its previous kernel finishes.

\subsection{Expert Weights and Cache}
\label{sec:cache_size}

The CPU branch reads expert weights from host memory, while the GPU branch uses cached or transferred weights. Since CPU expert execution is often limited by host-memory bandwidth, \myModel{} uses huge pages to improve memory access efficiency. The loader places the expert weights of each MoE layer from the GGUF file into a 2-MiB-aligned memory region, enables transparent huge pages, and registers the region with CUDA for GPU access. Huge pages improve CPU kernel performance by 6--7\% at large batch sizes.

On the GPU, a larger expert cache reduces weight transfers, so \myModel{} uses as much available GPU memory as possible. Each cache slot holds one expert per MoE layer, requiring 1.54~GB across DeepSeek-R1 Q4\_K\_M. To determine the cache capacity, \myModel{} estimates peak GPU memory usage from non-expert weights and buffers, expert cache slots, the batch KV cache, and the larger of the prefill and decode working sets. It allocates as many slots as this estimate allows. Since prefill generally requires more working memory than decoding, the cache starts at a prefill-safe size, expands after prefill to use the released memory, and shrinks again before the next prefill.

\subsection{AVX-512 CPU Expert Kernel}
\label{sec:cpu_kernels}

\begin{figure}[t]
  \centering
  \includegraphics[width=\columnwidth]{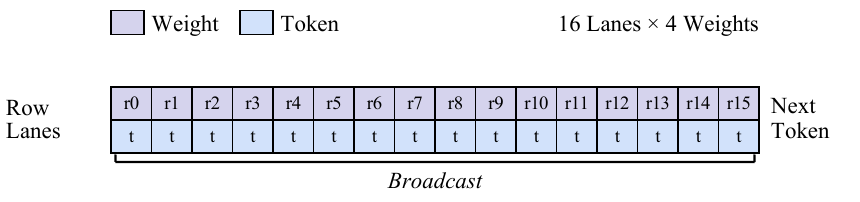}
  \caption{Lane layout of the AVX-512 CPU kernel. Each lane holds 4 weights of one row, the 16 lanes cover 16 rows, and one token is broadcast to all of them.}
  \label{fig:cpu_kernels}
  \vspace{-2em}
\end{figure}

The fused CPU path of KTransformers uses llamafile kernels to compute each expert as a row-major GEMM, processing tokens in groups of eight. For each group, they unpack the quantized weight blocks again, make a second pass over all rows for the remaining tokens, and apply the per-block quantization minimum separately for each row, block, and token. With 11--19 tokens per expert at large batches, this repeated work makes the kernels compute-bound.

\myModel{} replaces it with an AVX-512 VNNI kernel that keeps the same quantized inputs and integer dot products but changes the data layout and the loop order, so that every vector lane does useful work. As Figure~\ref{fig:cpu_kernels} shows, the kernel places weight rows in the lanes. It unpacks the quantized weights of 16 rows at a time into bytes and transposes them, so that each 32-bit lane of a 512-bit register holds 4 weights of one row. The \texttt{vpdpbusd} instruction multiplies them with the 4 input bytes of one token through an embedded broadcast, so the kernel needs neither a separate broadcast instruction nor padding when the tokens of an expert do not fill a vector. It also folds the offset of the quantized format into the initial integer accumulator.

For CPU experts in real DeepSeek-R1 inference calls with Q4\_K\_M weights on the H100 server, our kernel achieves 1.00$\times$ and 1.03$\times$ the performance of KTransformers during AIME2024 and MTBench decoding at batch sizes of 30 and 80, respectively. The speedups increase to 1.18$\times$ and 1.52$\times$ for ChatGPT-prompts and ShareGPT-V3 at batch sizes of 360 and 600, and reach 2.20--2.33$\times$ during prefill across all four datasets.

The only numerical difference from the KTransformers kernel is the order of floating-point summation, which changes the CPU branch output by a relative L2 error below $3.5\times10^{-4}$. \myModel{} keeps the original kernel on CPUs without AVX-512, such as the Zen 3 CPU of our A6000 server.

\subsection{Decode Loop and Host Runtime}
\label{sec:decode_loop}

Offline batches decode hundreds of requests together. If the decode loop processes them one request at a time, applying the logits warper and synchronizing the GPU with the host for each request incurs about 190~ms per decoding step for DeepSeek-R1 at batch size 600. \myModel{} eliminates this per-request overhead by processing logits in chunks of 64 rows to limit temporary memory usage. For each chunk, it applies temperature, top-p, and top-k in the same order as the original warper, producing bitwise-identical scores, and samples one token per request with a single batched multinomial operation. It also vectorizes the bookkeeping of all requests to eliminate per-request synchronizations.

\section{Evaluation}

\subsection{Experimental Setup}

\subsubsection{Platforms}

The experiments are conducted on two servers, (i) an AMD Ryzen Threadripper PRO 5995WX (64 cores, 512~GB host memory) with an NVIDIA RTX A6000 (48~GB) connected via PCIe 4.0 $\times$16, and (ii) an AMD EPYC 9654 (96 cores, 1{,}152~GB host memory) with an NVIDIA H100 (80~GB) connected via PCIe 5.0 $\times$16. Both use GCC 11.4, CUDA 12.6, and PyTorch 2.6.

\subsubsection{Models}

We evaluate DeepSeek-R1~\cite{deepseek-r1} and Kimi-K2~\cite{kimik2}, two frontier-scale MoE models. DeepSeek-R1 is a 61-layer model with 671B parameters and 58 MoE layers, each containing 256 routed experts and selecting 8 experts per token. Kimi-K2 is a 1T-parameter, 61-layer model with 60 MoE layers, each containing 384 routed experts and using top-8 routing. Both use Q4\_K\_M quantization~\cite{k_quant}, as native BF16 weights require approximately 1.34~TB and 2~TB, respectively, exceeding our host-memory capacity.

\subsubsection{Datasets}

We evaluate four datasets with diverse prompt types and scales, namely AIME2024~\cite{aime2024_dataset}, MTBench~\cite{mtbench}, ChatGPT-prompts~\cite{chatgpt_prompts_dataset}, and ShareGPT-V3~\cite{sharegpt_v3}. AIME2024 contains mathematical problems, while the other three cover reasoning and conversational prompts. For AIME2024, MTBench, and ChatGPT-prompts, we use all 30, 80, and 360 samples, respectively, as a single inference batch. From the 94{,}000 ShareGPT-V3 samples, we select prompts shorter than 256 tokens and form batches of 600. These batch sizes from 30 to 600 represent light to heavy expert activation workloads.

\begin{figure*}[t]
  \centering
  \includegraphics[width=1\linewidth]{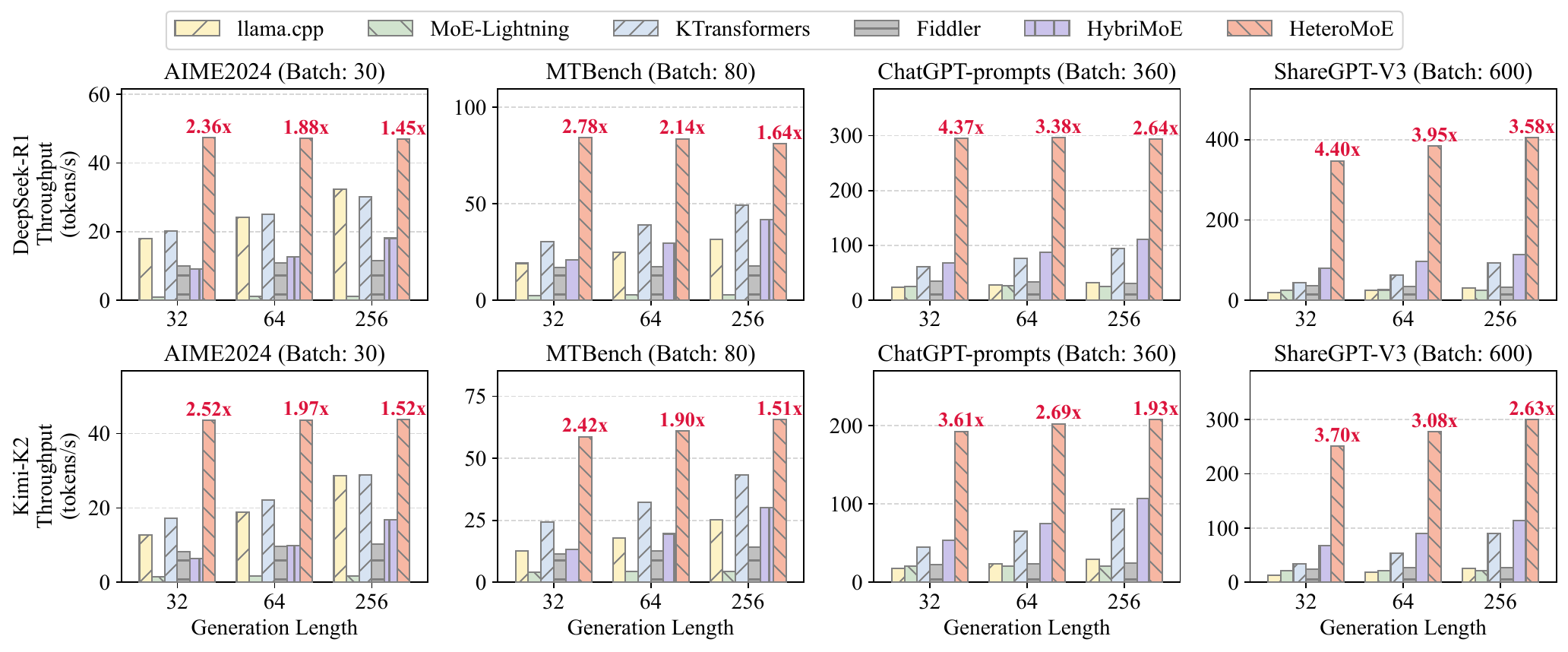}
  \caption{End-to-end throughput of DeepSeek-R1 and Kimi-K2 on H100. Labels give the speedup of \myModel{} over the strongest baseline.}
  \label{fig:e2e_throughput_h100}
  \vspace{-0.5em}
\end{figure*}

\begin{figure*}[t]
  \centering
  \includegraphics[width=1\linewidth]{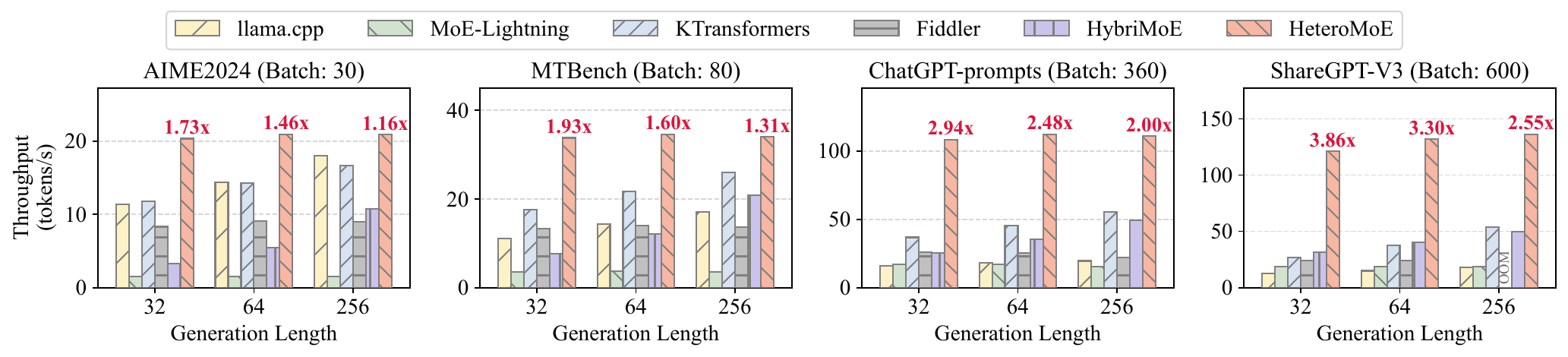}
  \caption{End-to-end throughput of DeepSeek-R1 across datasets on the A6000 platform. Labels give the speedup of \myModel{} over the strongest baseline.}
  \label{fig:e2e_throughput}
  \vspace{-1em}
\end{figure*}

\subsubsection{Baselines}
\label{sec:baselines}

We compare with five representative systems.

\begin{itemize}[leftmargin=*,topsep=0.1em]
\item \textbf{llama.cpp}~\cite{llamacpp} performs layer-wise CPU--GPU offloading. We keep all non-expert layers on the GPU, offload the experts of most MoE layers to the CPU, sweep its parallel slots, and report the best throughput for each workload.
\item \textbf{MoE-Lightning}~\cite{MoE-Lightning} keeps expert weights in host memory and streams them to the GPU for execution. Its CGOPipe pipeline overlaps CPU attention, GPU computation, and weight transfers across micro-batches, with the KV cache in host memory, and its hierarchical roofline model, HRM, selects the micro-batches and the GPU-resident experts for each workload.
\item \textbf{KTransformers}~\cite{ktransformers2024} runs attention and shared experts on the GPU and, following its original design, runs all routed experts on the CPU.
\item \textbf{Fiddler}~\cite{Fiddler} assigns active experts to the CPU or GPU using an online per-expert cost model.
\item \textbf{HybriMoE}~\cite{HybriMoE} also partitions active experts online between the CPU and GPU using per-expert costs, and caches hot experts on the GPU.
\end{itemize}

We calibrate the cost models of Fiddler and HybriMoE on each platform with measured single-expert CPU, GPU, and transfer times, as their papers do. Fiddler applies its rule in both prefill and decoding, and HybriMoE runs prefill experts on the GPU, as in their official code. All systems employ non-padding execution for variable-length batches, and utilize all CPU threads.

\subsubsection{Metrics and Parameter Settings}
\label{sec:metrics}

Our primary performance metric is \textbf{end-to-end throughput}, defined as the number of generated tokens divided by the total prefill and decoding time. Unless otherwise noted, each request generates exactly 32, 64, or 256 tokens. Decoding samples tokens with each model's default generation configuration and, except in the end-of-sequence experiment of Section~\ref{sec:batch_length}, ignores end-of-sequence tokens. We report the median of the runs of each configuration.

\subsection{End-to-End Performance}
\label{sec:e2e_throughput}

Figures~\ref{fig:e2e_throughput_h100} and~\ref{fig:e2e_throughput} report end-to-end throughput, with the improvement over the strongest other system annotated. \myModel{} is fastest in every setting. The strongest baseline changes across workloads, which shows why the CPU--GPU partition must follow the expert activation of each workload.

\subsubsection{H100 Platform with DeepSeek-R1}

\myModel{} improves the throughput over the strongest baseline by 1.45 to 4.40$\times$. At batch sizes 30 and 80, KTransformers is the strongest baseline at 32 and 64 tokens, and \myModel{} is 1.88 to 2.78$\times$ faster. Two effects add up. \myModel{} prefills in 3.3 and 5.0~s, against 22.6 and 41.9~s for the CPU experts of KTransformers, and it decodes 1.48 and 1.67$\times$ faster at 32 tokens by moving the hottest experts to the GPU. At batch sizes 360 and 600, HybriMoE is the strongest baseline and \myModel{} is 2.64 to 4.40$\times$ faster. Here \myModel{} decodes 2.87 and 3.80$\times$ faster than HybriMoE at 32 tokens and prefills in 6.4 and 16.0~s against 76.9 and 94.5~s.

Even at a generation length of 256 tokens, where decoding dominates, \myModel{} maintains speedups of 1.45--3.58$\times$ over the baselines. On AIME2024, it achieves 46.89~tokens/s compared with 32.35~tokens/s for llama.cpp, the strongest baseline. Its decoding throughput also exceeds that of every baseline by at least 37\%, demonstrating the performance advantage beyond prefill.

\subsubsection{A6000 Platform with DeepSeek-R1}

On the A6000 server, \myModel{} improves throughput over the strongest baseline by 1.16 to 3.86$\times$. The gains are smaller at small batches because this server lacks AVX-512 for the new CPU kernels, and its PCIe 4.0 link and slower host memory leave less room between the branches. At batch sizes 30 and 80, \myModel{} is 1.73 and 1.93$\times$ faster than KTransformers at 32 tokens and 1.16 and 1.31$\times$ faster than the strongest baseline at 256 tokens. At batch sizes 360 and 600, it is 2.00 to 3.86$\times$ faster.

\subsubsection{H100 Platform with Kimi-K2}

The bottom row of Figure~\ref{fig:e2e_throughput_h100} evaluates a larger model with 384 experts per layer. \myModel{} improves throughput over the strongest baseline by 1.51 to 3.70$\times$, reaching 43.6 to 250.9~tokens/s at 32 tokens and 43.7 to 299.9~tokens/s at 256 tokens. As for DeepSeek-R1, KTransformers is the strongest baseline at batch sizes 30 and 80, and HybriMoE at 360 and 600. \myModel{} therefore scales to a trillion-parameter model with more experts per layer.

\subsubsection{Branch Times}
We measure the time that each branch spends on the routed experts of a decoding MoE call on the H100 server with 256 generated tokens. On MTBench at batch size 80, KTransformers runs all routed experts on the CPU, which takes 18.3~ms per call, whereas the CPU and GPU+PCIe branches of \myModel{} take 12.6 and 12.8~ms. On ShareGPT-V3 at batch size 600, the two branches take 52.3 and 3.2~ms with HybriMoE and 18.2 and 19.5~ms with \myModel{}. On Kimi-K2 with AIME2024 at batch size 30, the CPU takes 12.2~ms with KTransformers, while the two branches of \myModel{} take 8.0 and 8.7~ms. \myModel{} thus balances the two branches while the baselines leave one of them nearly idle, and its decoding MoE time is 1.35, 2.65, and 1.34$\times$ shorter. Section~\ref{sec:policy_comparison} isolates the effect of the partition on one backend.

\subsection{Cost Model Accuracy}
\label{sec:model_accuracy}

\begin{table}[t]
\centering
\caption{Signed error of the predicted times on end-to-end runs, relative to the measured times. Negative values are underestimates.}
\label{tab:model_accuracy}
\begin{adjustbox}{max width=\columnwidth}
\begin{tabular}{ll|ccc}
\toprule
Server & Phase & CPU & GPU+PCIe & Call \\
\midrule
\multirow{2}{*}{H100} & Decoding & $+2.6\%$ & $-2.8\%$ & $-1.8\%$ \\
 & Prefill & $-0.1\%$ & $-4.6\%$ & $-5.5\%$ \\
\midrule
\multirow{2}{*}{A6000} & Decoding & $-0.7\%$ & $-5.7\%$ & $-3.8\%$ \\
 & Prefill & $+6.7\%$ & $-0.3\%$ & $-0.9\%$ \\
\bottomrule
\end{tabular}
\end{adjustbox}
\vspace{-1em}
\end{table}

The scheduler chooses each partition from the predicted branch times, so the cost model must track the measured times across workloads. We evaluate it on held-out end-to-end runs that the calibration fit never sees, ChatGPT-prompts at batch size 360 and ShareGPT-V3 at batch size 600 with 256 generated tokens. For every MoE call, we feed the logged expert activation, cached experts, and number of transferred experts to the cost model with the constants of the calibration procedure in Section~\ref{sec:online_partitioning}. We then compare its predictions with the measured times of the CPU branch, the GPU+PCIe branch, and the call, whose time is the longer of the two branches. Table~\ref{tab:model_accuracy} reports the median signed errors, excluding the two warm-up prefill calls of each run. On the H100 server, the predicted decoding call is 1.8\% shorter than the measured one, and prefill calls are predicted 5.5\% short. On the A6000 server, the model underestimates decoding calls by 3.8\%, mostly in the GPU+PCIe branch, and prefill calls by 0.9\%.

\subsection{Same-Backend Policy Comparison}
\label{sec:policy_comparison}

\begin{table}[t]
\centering
\caption{Throughput (tokens/s) and GPU share of partitioning policies on one backend. All policies run inside \myModel{} on A6000 with 256 generated tokens.}
\label{tab:vs_diff_strategy}
\begin{adjustbox}{max width=\columnwidth}
\begin{tabular}{l|cc|cc}
\toprule
\multirow{2}{*}{Partitioning policy} & \multicolumn{2}{c|}{ChatGPT-prompts, 360} & \multicolumn{2}{c}{ShareGPT-V3, 600} \\
 & tokens/s & GPU share & tokens/s & GPU share \\
\midrule
Fiddler rule & 60.76 & 38.0\% & 51.89 & 72.1\% \\
HybriMoE rule & 63.09 & 34.9\% & 70.18 & 49.6\% \\
\myModel{} cost model & \textbf{113.77} & 17.9\% & \textbf{140.04} & 21.2\% \\
\midrule
Theoretical best & 131.2 & -- & 159.3 & -- \\
\bottomrule
\end{tabular}
\end{adjustbox}
\vspace{-1em}
\end{table}

To isolate the partition policy, we replace only the per-call decision inside the A6000 configuration of \myModel{} and keep its kernels, cache, constants, and pipeline. All runs generate 256 tokens. The \textbf{Fiddler rule} runs cached experts on the GPU and any other expert on the GPU when its single-expert CPU time exceeds the GPU and transfer time. The \textbf{HybriMoE rule} is a line-by-line port of its hybrid scheduling, which fills the GPU queue from the most loaded experts and the CPU queue from the least loaded ones until the two queues meet. Both rules use per-expert CPU costs measured with our own CPU kernel at startup.

Table~\ref{tab:vs_diff_strategy} shows that the additive rules reduce throughput by assigning 34.9--72.1\% of active experts to the GPU, compared with 17.9--21.2\% for our cost model. This increases expert transfers from 36.3--53.6 to 78.0--182.7 per decoding call. At 1.18~ms per transfer, the GPU and PCIe branch becomes the critical path. The excessive offloading stems from summing single-expert costs, which overestimates the fused CPU call by 4.0--4.8$\times$. Our cost model thus outperforms Fiddler by 1.87 and 2.70$\times$ and HybriMoE by 1.80 and 2.00$\times$ on ChatGPT-prompts and ShareGPT-V3, respectively. We also compare against the theoretical best, which assumes the best CPU--GPU partition, an ideal expert cache, and the highest rates measured on this server. Our cost model achieves 87\% and 88\% of the theoretical best, with the remaining gap mainly due to expert execution below these rates rather than the partition or cache.

Instrumented runs of the same settings show where the additive rules lose time. With the Fiddler rule, the median CPU and GPU+PCIe branch times per decoding call are 38.3 and 85.3~ms on ChatGPT-prompts and 15.6 and 196.8~ms on ShareGPT-V3, and with the HybriMoE rule they are 31.8 and 101.8~ms and 35.5 and 131.2~ms. Expert transfers take more than 99\% of these GPU+PCIe times, so the additive rules overload the GPU+PCIe branch while the CPU branch finishes early. \myModel{} instead balances the two branches, at 45.2 and 50.2~ms on ChatGPT-prompts and 63.2 and 57.5~ms on ShareGPT-V3.

We further validate our hottest-first expert selection by replaying measured DeepSeek-R1 routing on the H100 server using our cost model. For each dataset, we sample 16 decoding MoE calls across layers and steps and 16 prefill MoE layers. For each sample, we fix the number of transfers chosen by the scheduler and compare the hottest uncached experts against alternative sets of the same size. These include 128 random subsets, the coldest experts, experts ordered by ID regardless of activation, and near-hottest sets formed by swapping one of the last eight selected experts with one of the next eight. We also perform a local search that starts from the hottest set and repeatedly swaps a transferred expert with a CPU expert whenever the modeled time decreases. The best alternative reduces modeled MoE time by at most 0.12\% for decoding and 0.29\% for prefill, both below the cost model's prediction error. In contrast, random, ID-ordered, and coldest-first selections increase modeled decoding time by 2--13\% at batch sizes of 360 and 600.

\subsection{Batch Size and Generation Length}
\label{sec:batch_length}

\subsubsection{Batch Size}

\begin{figure}[t]
  \centering
  \includegraphics[width=1\linewidth]{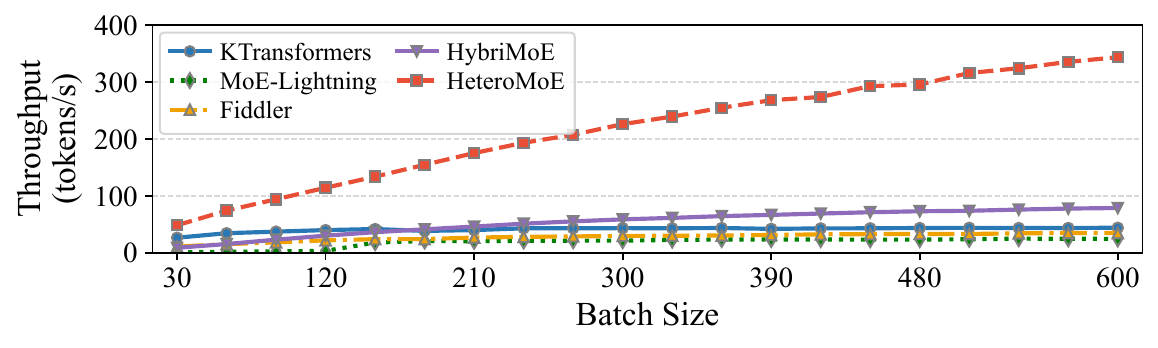}
  \caption{Throughput sensitivity to batch size for ShareGPT-V3 prompts on the H100 server at 32 generated tokens.}
  \label{fig:througput_varing_batch}
  \vspace{-1em}
\end{figure}

Figure~\ref{fig:througput_varing_batch} varies the batch size of ShareGPT-V3 prompts from 30 to 600 at 32 generated tokens on the H100 server. KTransformers is competitive at small batches but saturates early, at 40 to 44~tokens/s from batch size 120 on. HybriMoE benefits from concurrent CPU--GPU execution at larger batches and is the strongest baseline from batch size 180 on, while Fiddler stays below it. MoE-Lightning jumps from 3.51 to 18.10~tokens/s between batch sizes 120 and 150, where its HRM policy changes the micro-batch size. \myModel{} achieves the highest throughput at every batch size. Its speedup over the strongest baseline is 1.82$\times$ at batch size 30, rises to 3.75 to 3.84$\times$ at batch sizes 240 to 300, and reaches 4.36$\times$ at 600, because the scheduler spreads the growing activation over both branches. Figure~\ref{fig:batch_motivation} normalizes this sweep to a theoretical best, which gives every routed-expert MoE call its roofline lower bound under the best CPU--GPU partition and keeps the measured time of all other work. This bound uses the highest rates measured on this server for host-memory bandwidth, PCIe bandwidth, and CPU and GPU compute, and an ideal GPU expert cache as large as that of \myModel{}. \myModel{} reaches 82 to 87\% of the theoretical best from batch 60 on.

\subsubsection{Generation Length}

\begin{figure}[t]
  \centering
  \includegraphics[width=1\linewidth]{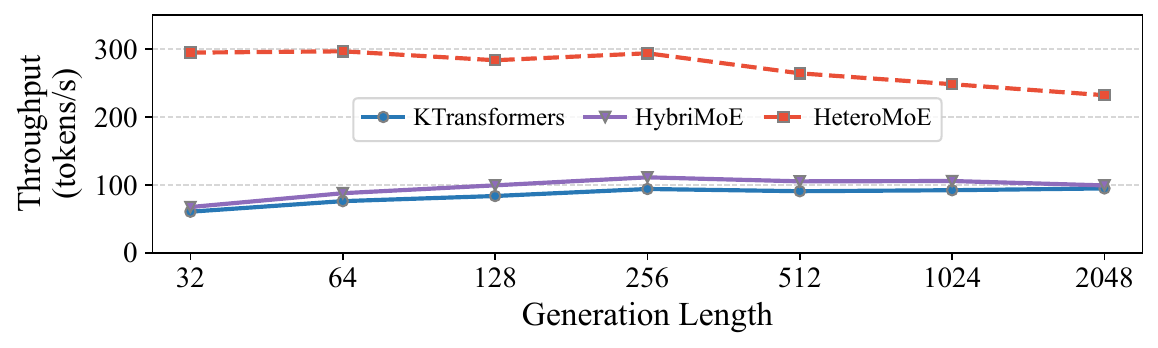}
  \caption{Throughput across generation lengths for ChatGPT-prompts at batch size 360 on the H100 server.}
  \label{fig:througput_varing_length}
  \vspace{-1em}
\end{figure}

Figure~\ref{fig:througput_varing_length} varies the generation length of ChatGPT-prompts at batch size 360 from 32 to 2{,}048 tokens on the H100 server. \myModel{} stays the fastest at every length. Its throughput falls from 295 to 232~tokens/s, because the KV cache of longer sequences leaves fewer GPU slots for experts, while KTransformers and HybriMoE stay between 61 and 111~tokens/s. Its speedup over the strongest baseline therefore decreases from 4.37$\times$ at 32 tokens to 2.34$\times$ at 2{,}048. With 4{,}096 generated tokens on AIME2024 at batch size 180, \myModel{} reaches 134.6~tokens/s, 1.64$\times$ that of KTransformers.

\subsubsection{End-of-Sequence Stops}
The runs above keep every request in the batch until the last step. We also run DeepSeek-R1 on the H100 server with end-of-sequence handling, where a request leaves the batch at its first end-of-sequence token or at a length cap, and compare \myModel{} with KTransformers and HybriMoE under the same sampling configuration and cap. On MTBench at batch size 80 with a cap of 2{,}048 tokens, \myModel{} generates 62.68~tokens/s against 43.34 for KTransformers and 30.89 for HybriMoE, 1.45$\times$ the strongest baseline. On ChatGPT-prompts at batch size 360 with a cap of 1{,}024 tokens, it generates 254.91~tokens/s against 100.42 and 111.03, 2.30$\times$ the strongest baseline. These margins stay close to the fixed-length ones, 1.64$\times$ at 256 tokens on MTBench and 2.35$\times$ at 1{,}024 tokens on ChatGPT-prompts, although about 70\% of the MTBench requests and 30\% of the ChatGPT-prompts requests end early and the batch shrinks accordingly.

\subsection{System Component Ablation}
\label{sec:ablation}

\begin{figure}[t]
  \centering
  \includegraphics[width=1\linewidth]{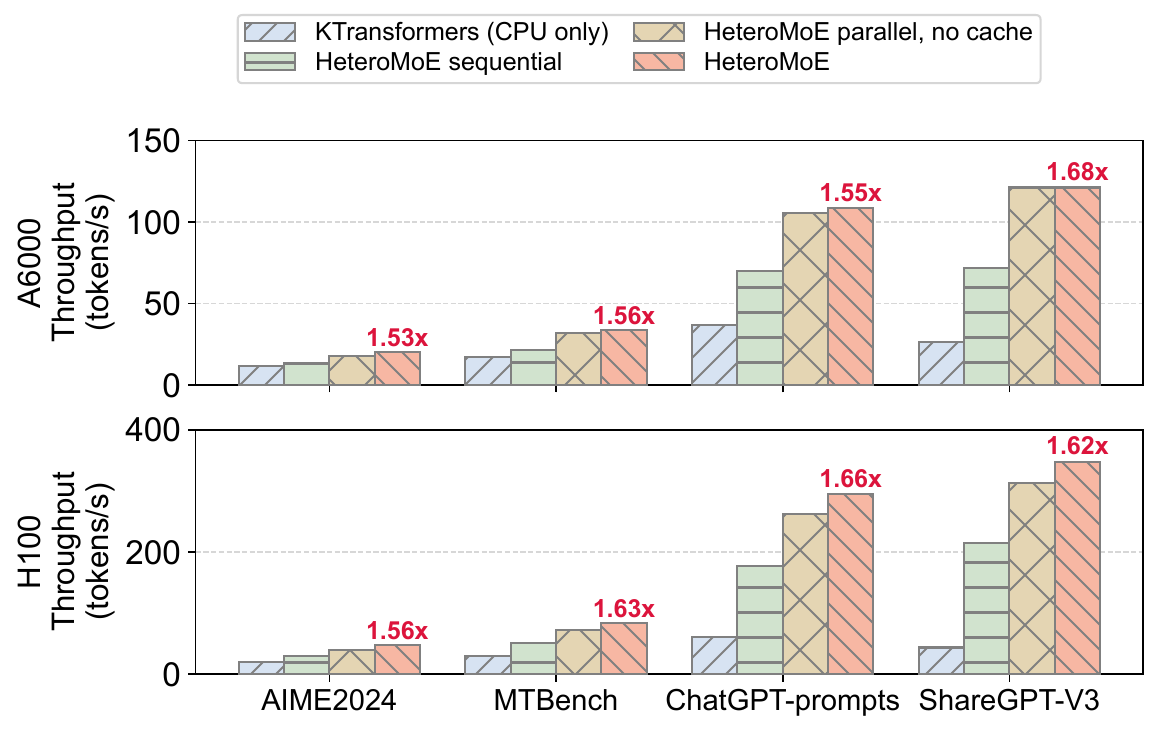}
  \caption{Throughput of CPU-only, sequential, uncached, and full execution on the A6000 and H100 servers. Labels give the speedup of full over sequential execution.}
  \label{fig:ablation}
  \vspace{-1em}
\end{figure}

The partition determines where experts run, while the execution engine determines how efficiently they execute. Figure~\ref{fig:ablation} evaluates the engine optimizations on the A6000 and H100 servers with 32-token generation. The \textbf{sequential} variant executes the CPU and GPU branches sequentially while retaining the same partition, kernels, and cache. The \textbf{no-cache} variant executes both branches concurrently but disables the GPU expert cache.

\subsubsection{CPU--GPU Parallelism}
On the A6000 server, parallel execution improves throughput by 1.53, 1.56, 1.55, and 1.68$\times$ across the four workloads, with decoding speedups of 1.48--1.77$\times$. The gains generally increase with batch size as both branches become longer. On the H100 server, parallel execution improves throughput by 1.56, 1.63, 1.66, and 1.62$\times$ across the four workloads, with decoding speedups of 1.55--1.70$\times$. Even without parallelism, sequential hybrid execution outperforms CPU-only KTransformers by 1.13--2.70$\times$ on the A6000 server and 1.51--4.90$\times$ on the H100 server, as running the hottest experts on the GPU speeds up both phases.

\subsubsection{GPU Expert Caching}
On the A6000 server, caching 17, 16, and 11 experts per layer improves throughput by 12\%, 6\%, and 3\% on AIME2024, MTBench, and ChatGPT-prompts, respectively. On the H100 server, caching 34--43 experts per layer improves throughput by 21\%, 17\%, 13\%, and 11\% across the four workloads. The benefit decreases with batch size because more experts become active, and the cache covers a smaller fraction of them.

\subsection{Overhead}
\label{sec:overhead}

\subsubsection{Online Decision}

The partition decision runs on the host at every MoE call. On the A6000 server with ShareGPT-V3 prompts, the decision of a decoding MoE call, including counting, sorting, both branch models, and the bookkeeping of the CPU and GPU sets, takes 0.10 to 0.17~ms of host time across batch sizes 30 to 600. Its share of the MoE call falls from 0.52\% at batch size 30 to 0.25\% at 600 as expert computation grows. The time stays nearly flat across batch sizes, because the cost evaluation depends on the number of active experts, at most 256, and counting the routed IDs is a single vector operation. For prefill calls of 4{,}244 to 37{,}915 tokens, the decision takes 0.25 to 0.58~ms, under 0.2\% of the call.

\subsubsection{Calibration}

The calibration runs of Section~\ref{sec:online_partitioning} take about 6~min on the H100 server and 15~min on the A6000 server, and the least-squares fit takes about 2~s. Constants from these runs give the same speed as the constants used in our evaluation, within 1\%. Constants calibrated at one batch size carry over to similar batch sizes but not across the whole batch range, so the calibration runs cover small and large batches as well as short and long prompts.

\subsection{Quality Validation}
\label{sec:quality}

We evaluate the impact of dynamic CPU--GPU execution on task quality on H100. For AIME2024, we sample each question five times with temperature 0.6, top-p 0.95, and a 10K-token limit. For 40 objective MTBench questions, we run each system twice with an 8K-token limit. Pass@k measures the number of questions solved within k samples.

\begin{table}[t]
\centering
\caption{Quality comparison on DeepSeek-R1 Q4\_K\_M, with pass@1 as mean $\pm$ std over runs.}
\label{tab:quality_validation}
\begin{adjustbox}{max width=1\columnwidth}
\begin{tabular}{ll|ccc}
\hline
Dataset & Metric & KTransformers & \myModel{} & HybriMoE \\ \hline
AIME2024 & pass@1 & 57.3$\pm$2.8\% & 57.3$\pm$3.7\% & 63.3$\pm$5.3\% \\
AIME2024 & pass@5 & 63.3\% & 73.3\% & 80.0\% \\
MTBench & pass@1 & 95.0$\pm$0.0\% & 95.0$\pm$3.5\% & 92.5$\pm$3.5\% \\
MTBench & pass@2 & 95.0\% & 97.5\% & 97.5\% \\
\hline
\end{tabular}
\end{adjustbox}
\end{table}

Table~\ref{tab:quality_validation} shows no quality degradation from dynamic partitioning. On AIME2024, \myModel{} and KTransformers have the same pass@1, and the pass@5 difference corresponds to three of the 30 questions. On the 40 MTBench objective questions, their pass@1 is also identical and the pass@2 difference corresponds to one question. \myModel{} preserves all routed token-expert computations and weights, changing only their execution location and order. Because decoding uses nonzero-temperature sampling and heterogeneous CPU--GPU floating-point accumulation can produce different autoregressive outputs, every system varies from run to run, with pass@1 standard deviations of up to 5.3 points on AIME2024 and 3.5 points on MTBench, and differences of this magnitude are within this variation. The official GPU kernels of HybriMoE use the Marlin int4 format, so it computes its GPU experts with Marlin int4 weights derived from Q4\_K\_M and its CPU experts with Q4\_K\_M, as the other two systems do. Its outputs therefore differ through the weight format of these GPU experts as well as through sampling.

\section{Related Work}

\subsection{Memory-Constrained GPU-only Inference}
Under limited GPU memory, GPU-only systems keep model tensors in host memory or on disk but execute model operators on the GPU. Klotski~\cite{Klotski} reuses loaded expert weights across multiple batches and reorders expert execution to overlap the remaining transfers. Diff-MoE~\cite{Diff-MoE} improves expert reuse through priority-driven multi-level caching. MoE-Infinity~\cite{MoE-Infinity}, ProMoE~\cite{promoe}, and fMoE~\cite{fmoe} further reduce on-demand transfers through activation-aware caching, proactive prefetching, and fine-grained expert offloading. Although these techniques reduce transfers through reuse and prediction, the CPU serves only as a storage tier, and performance still depends on balancing PCIe transfers against GPU computation.

\begin{table}[t]
\centering
\caption{Heterogeneous expert execution designs.}
\label{tab:vs_baseline}
\begin{adjustbox}{max width=\columnwidth}
\begin{tabular}{l|ccc}
\toprule
System & \begin{tabular}[c]{@{}c@{}}Expert\\comp.\end{tabular} & \begin{tabular}[c]{@{}c@{}}CPU cost\\model\end{tabular} & \begin{tabular}[c]{@{}c@{}}Partition\\outcome\end{tabular} \\
\midrule
MoE-Lightning~\cite{MoE-Lightning} & GPU & -- & Regime-specific \\
Klotski~\cite{Klotski} & GPU & -- & Regime-specific \\
MoE-Gen~\cite{MOE-Gen} & GPU & -- & Regime-specific \\
MoE-Lens~\cite{MOE-Lens} & GPU & -- & Regime-specific \\
Diff-MoE~\cite{Diff-MoE} & GPU & -- & Regime-specific \\
\midrule
KTransformers~\cite{ktransformers2024} & CPU & -- & Regime-specific \\
\midrule
Fiddler~\cite{Fiddler} & CPU + GPU & Per-expert sum & GPU-heavy \\
HybriMoE~\cite{HybriMoE} & CPU + GPU & Per-expert sum & GPU-heavy \\
MoE-Balance~\cite{moebalance} & CPU + GPU & Per-expert sum & GPU-heavy \\
CoX-MoE~\cite{coxmoe} & CPU + GPU & -- & Regime-specific \\
\midrule
\myModel{} & CPU + GPU & Fused roofline & Balanced \\
\bottomrule
\end{tabular}
\end{adjustbox}
\vspace{-1em}
\end{table}

\subsection{Heterogeneous CPU--GPU Inference}
Heterogeneous systems additionally execute selected model operators on the CPU. For dense LLMs, FlexGen~\cite{FlexGen} can delegate decoding attention to the CPU when the KV cache resides in host memory, while FastDecode~\cite{he2024fastdecode} disaggregates attention computation to CPUs. HeteGen~\cite{HeteGen} and LiA~\cite{LIA} coordinate CPU and GPU execution at a finer granularity, and PowerInfer~\cite{powerinfer,powerinfer2} computes hot and cold activated neurons on the GPU and CPU, respectively. For MoE models, MoE-Lightning~\cite{MoE-Lightning} and MoE-Lens~\cite{MOE-Lens} place attention on the CPU and expert GEMMs on the GPU, and MoE-Gen~\cite{MOE-Gen} similarly schedules attention across both devices while batching expert computation on the GPU. KTransformers~\cite{ktransformers2024} instead executes routed experts on the CPU, llama.cpp~\cite{llamacpp} partitions the model by layer, and Fiddler~\cite{Fiddler}, HybriMoE~\cite{HybriMoE}, MoE-Balance~\cite{moebalance}, and Huang et al.~\cite{taiwan_efficientcpugpucollaborativeinference} partition expert computation between the CPU and GPU. CoX-MoE~\cite{coxmoe} runs CPU experts with AMX and places frequently activated experts on the GPU ahead of time. Table~\ref{tab:vs_baseline} summarizes these designs. The fixed placements of the GPU-only systems, KTransformers, llama.cpp, and CoX-MoE suit the activation of particular batch regimes. Their partitioning, however, sums fixed per-expert costs, so it misses the shifting limits of the fused CPU call and transfers too many experts. \myModel{} instead estimates the fused CPU call with a roofline and balances it against the expert transfers at every call.

\section{Conclusion}

We presented \myModel{}, an activation-aware CPU--GPU expert scheduling system for high-throughput MoE inference on memory-constrained servers. At each MoE call, a calibrated cost model dynamically partitions experts between the CPU and GPU, accounting for expert activation, computation, PCIe transfers, and GPU caching. An optimized execution engine overlaps CPU and GPU computation, prefetches experts, and minimizes synchronization and decode-loop overhead. Evaluations on DeepSeek-R1 and Kimi-K2 across two hardware platforms show that \myModel{} achieves up to 4.40$\times$ higher end-to-end throughput than existing systems.

\bibliographystyle{ACM-Reference-Format}
\bibliography{refs}

\end{document}